\documentclass[journal,draftclsnofoot,onecolumn]{IEEEtran}

\usepackage{setspace}
\usepackage{amsmath}
\usepackage{amssymb}
\usepackage{mathrsfs}
\usepackage{amsthm}
\usepackage{multirow}
\usepackage{color}
\usepackage{longtable}
\usepackage{array}
\usepackage{url}
\usepackage{comment}
\usepackage{enumerate}
\usepackage{eucal}
\usepackage{graphics}

\usepackage{algorithm}
\usepackage{algpseudocode}
\usepackage[noadjust]{cite}
\usepackage[table]{xcolor}

\usepackage{stmaryrd}

\usepackage{mathtools}
\usepackage{xparse}
\DeclarePairedDelimiterX{\set}[1]{\{}{\}}{\setargs{#1}}
\NewDocumentCommand{\setargs}{>{\SplitArgument{1}{;}}m}
{\setargsaux#1}
\NewDocumentCommand{\setargsaux}{mm}
{\IfNoValueTF{#2}{#1} {#1\,\delimsize|\,\mathopen{}#2}}

\DeclarePairedDelimiter\abs{\lvert}{\rvert}
\DeclarePairedDelimiter\ceil{\lceil}{\rceil}

\DeclarePairedDelimiter\parenv{\lparen}{\rparen}

\newcommand{\cB}{\mathcal{B}}
\newcommand{\cC}{\mathcal{C}}
\newcommand{\cD}{\mathcal{D}}
\newcommand{\cE}{\mathcal{E}}

\newcommand{\cL}{\mathcal{L}}
\newcommand{\cM}{\mathcal{M}}
\newcommand{\cN}{\mathcal{N}}

\newcommand{\cS}{\mathcal{S}}

\newcommand{\cU}{\mathcal{U}}
\newcommand{\cV}{\mathcal{V}}

\newcommand{\cY}{\mathcal{Y}}
\newcommand{\cZ}{\mathcal{Z}}

\renewcommand{\leq}{\leqslant}

\renewcommand{\geq}{\geqslant}

\newcommand{\Maj}{\rm Maj}

\theoremstyle{plain}
\newtheorem{theorem}{Theorem}
\newtheorem{corollary}[theorem]{Corollary}
\newtheorem{lemma}[theorem]{Lemma}
\newtheorem{proposition}[theorem]{Proposition}

\theoremstyle{definition}
\newtheorem{definition}[theorem]{Definition}

\newtheorem*{remark}{Remark}

\newcommand{\Z}{\mathbb{Z}}
\newcommand{\N}{\mathbb{N}}

\newcommand{\ve}{\mathbf{e}}

\newcommand{\vs}{\mathbf{s}}
\newcommand{\vv}{\mathbf{v}}
\newcommand{\vu}{\mathbf{u}}
\newcommand{\vx}{\mathbf{x}}
\newcommand{\vy}{\mathbf{y}}
\newcommand{\vz}{\mathbf{z}}

\newcommand{\vc}{\mathbf{c}}

\DeclareMathOperator{\wt}{wt}

\newcommand{\kp}{k_+}
\newcommand{\km}{k_-}
\newcommand{\BALL}{{\mathcal B}(n,t,\kp,\km)}
\newcommand{\eqdef}{\triangleq}
\newcommand{\splt}{\diamond}

\begin{document}
\date{}
\title{Sequence Reconstruction for Limited-Magnitude Errors}
\author{
Hengjia Wei and Moshe~Schwartz,~\IEEEmembership{Senior Member,~IEEE}%
\thanks{Hengjia Wei is with the School
   of Electrical and Computer Engineering, Ben-Gurion University of the Negev,
   Beer Sheva 8410501, Israel
   (e-mail: hjwei05@gmail.com).}%
\thanks{Moshe Schwartz is with the School
   of Electrical and Computer Engineering, Ben-Gurion University of the Negev,
   Beer Sheva 8410501, Israel
   (e-mail: schwartz@ee.bgu.ac.il).}%
\thanks{This work was supported in part by the Israel Science Foundation (ISF) under grant No.~270/18.}
}
\maketitle

\begin{abstract}
    Motivated by applications to DNA storage, we study reconstruction and list-reconstruction schemes for integer vectors that suffer from limited-magnitude errors. We characterize the asymptotic size of the intersection of error balls in relation to the code's minimum distance. We also devise efficient reconstruction algorithms for various limited-magnitude error parameter ranges. We then extend these algorithms to the list-reconstruction scheme, and show the trade-off between the asymptotic list size and the number of required channel outputs. These results apply to all codes, without any assumptions on the code structure. Finally, we also study linear reconstruction codes with small intersection, as well as show a connection to list-reconstruction codes for the tandem-duplication channel.
\end{abstract}


\section{Introduction}
\IEEEPARstart{T}{he} \emph{sequence-reconstruction problem}, which was first introduced by Levenshtein \cite{Lev01a}, considers a paradigm in which a sequence from some set is transmitted multiple times over a channel and the receiver needs to recover the transmitted sequence from the  received sequences. It was originally motivated by the communication scenario where the only feasible strategy to combat errors is repeated transmission.  Recently, it has been observed that this problem has a natural connection to DNA-based data storage systems. In such systems,  the DNA strands are expected to be replicated many times, whether due to biological processes when in-vivo storage is used, or due to chemical processes in synthesis or sequencing when in-vitro storage is used. The user usually gets many noisy reads of the stored DNA strand when retrieving the data. Recovering the original DNA strand from its multiple noisy reads is therefore a sequence-reconstruction problem.

For a sequence $\vx$, the \emph{error ball} of $\vx$ is the set of all  possible outputs with $\vx$ being transmitted through the channel.  Clearly, the  number of different channel outputs required to recover the transmitted sequence must be larger than the maximum intersection between the error balls of any two possible transmitted sequences \cite{Lev01a}. One goal of the reconstruction  problem is to determine the maximum intersection of two balls where the distance between their centers is at least some prescribed value. A significant number of papers has been devoted to  determining this value for various error models, including substitutions, deletions, insertions, transpositions, and tandem-duplications \cite{Lev01,Lev01a,KonLevSie07,SalGabSchDol17,GabTaa18,YehSch20}. Additionally, a graph-theoretical approach was studied in \cite{Lev01a,LevKonKonMol08,LevSie09} to solve this problem in a more general metric distance.

Apart from determining the maximum intersection of two  balls, other research directions have been considered as well. \cite{YaaBru19,SinYaa19} proposed efficient reconstruction algorithms to combat substitutions. Reconstruction codes have been designed to  recover the transmitted sequences from a given number of  sequences corrupted by tandem-duplications \cite{YehSch20} or a single edit error \cite{CaiKiaNguYaa20}. Yaakobi and Bruck extended this problem in the context of associative memories \cite{YaaBru19} and introduced the notion of \emph{uncertainty} of an associative memory  for information retrieval, the value of which is equal to the maximum intersection of multiple error balls. In the context of sequence reconstruction, a closely related problem is   to construct a list from multiple received noisy sequences such that the  transmitted sequence  is included in the list. The  trade-off between the size of the minimum list (the number of  balls) and the number of different received noisy sequences (the maximum intersection / uncertainty) has been analyzed  for substitutions \cite{JunLaiLeh21} and for tandem-duplications \cite{YehSch21}.

This paper focuses  on \emph{limited-magnitude errors}, which  could  be found in several  applications, including high-density magnetic recording channels \cite{KuzVin93,LevVin93}, flash memories \cite{CasSchBohBru10}, and some DNA-based  storage systems \cite{JaiFarSchBru20,WeiSch21}. In all of these applications, information  is encoded as vectors of integers, and these vectors are affected by noise that may increase or decrease entries of the vectors by a limited amount. The design of codes combating such errors, or equivalently, the packing/tiling of the corresponding errors balls, has been extensively researched, see  e.g.,
\cite{Ste84,HamSte84,HicSte86,Ste90,SteSza94,KloLuoNayYar11,YarKloBos13,Sch14,ZhaGe16,ZhaZhaGe17,ZhaGe18,YeZhaZhaGe20,BuzEtz12,WeiSch21}, and the many
references therein.

In this paper, we study the reconstruction problem with respect to limited-magnitude errors.
We first propose a new kind of distance to capture the capability of correcting limited-magnitude errors. Then for any code $\cC$ of distance at least a prescribed value, we present both an upper bound and a lower bound on the size of the maximum intersection of any two error balls centered at the  codewords of $\cC$. In this way, we characterize the  trade-off between this value and the number of excessive errors that the code $\cC$ cannot cope with. Moreover, we study this reconstruction problem  in a group-theoretical approach. In the channel that introduces a single limited-magnitude error, we design two classes of reconstruction codes which  both have densities  significantly larger than that of the normal error-correcting codes, at the cost of requiring one or two more received sequences.  We present two efficient  algorithms to reconstruct a transmitted codeword from any given code.  Finally, we modify our reconstruction algorithms to accommodate the requirement of list decoding when the number of received sequences is less than the maximum intersection. Additionally, we show that one of our reconstruction algorithm could be used in the context of tandem duplications.

The paper is organized as follows.
Section~\ref{sec:prelim} provides notation and basic known
results used throughout the paper. In Section~\ref{sec:intersection} we study the maximum intersection and present a few upper bounds and lower bounds. In Section~\ref{sec:codedesign},  we design codes that can recover the sequence from two or three received sequences. Section~\ref{sec:algorithm} presents two efficient reconstruction algorithms. 
Section~\ref{sec:listreconstruction} studies the list decoding problem with multiple received sequences. Section~\ref{sec:tandemdup} discusses the reconstruction algorithm for tandem duplications.

\section{Preliminaries}
\label{sec:prelim}
Let $\Z$ denote the ring of integers and $\N$ denote the set of natural numbers. Throughout the paper we let $n$ and $t$ be integers such that $n\geq
t\geq 1$. We further assume $\kp$ and $\km$ are non-negative integers
such that $\kp\geq \km\geq 0$. For integers $a\leq b$ we define
$[a,b]\eqdef\set*{a,a+1,\dots,b}$ and $[a,b]^*\eqdef
[a,b]\setminus\set*{0}$. Vectors will be written using bold lower-case letters. If $\vv=(v_1,\dots,v_n)$ is a vector, we shall conveniently use $\vv[i]$ to denote its $i$-th entry, namely, $\vv[i]\eqdef v_i$.

For an integer vector $\vv\in\Z^n$, if  $t$ of its entries suffer
an increase by as much as $\kp$, or a decrease by as much as $\km$, we say $\vv$ suffers \emph{$t$  $(\kp,\km)$-limited-magnitude errors}.
We define the \emph{$(n,t,\kp,\km)$-error-ball} as
\begin{equation}
  \label{eq:ball}
  \BALL\triangleq \set*{\vx=(x_1,x_2,\ldots,x_n)\in \Z^n ; -\km \leq x_i \leq \kp \text{ and  } \wt(\vx) \leq t  },
\end{equation}
where $\wt(\vx)$ denotes the Hamming weight of $\vx$. Thus, the translate $\vv+\BALL$ is the error ball centered at the vector $\vv$. If the values of $\kp$ and $\km$ can be inferred from the context, we simply denote it as $\cB_t(\vv)$ to emphasize its center and  radius.

The size of the ball $\BALL$ will appear throughout this work. It is closely related to the size of a ball in the Hamming metric, which over an alphabet of size $q$, and with radius $t$, is
\[ V_q(n,r) \eqdef \sum_{i=0}^{t} \binom{n}{i}(q-1)^i.\]
Using this notation, the size of $\BALL$ is given by
\[ \abs*{\BALL}=V_{\kp+\km+1}(n,t).\]

An error-correcting code in this setting  is a packing of $\Z^n$ by $\BALL$, that is, a subset $\cC \subseteq \Z^n$ such that for any two distinct vectors $\vx,\vy\in\cC$,  the balls $\vx+\BALL$ and $\vy+\BALL$ are disjoint. The largest integer $t$ with this property is the \emph{unique-decoding radius of $\cC$} (with respect to $(\kp,\km)$-limited-magnitude errors).

Throughout this paper, we use these two notions of error-correcting code and packing interchangeably. The following distance\footnote{Note that although $d_{\ell}$ is referred to as a distance in \cite{CasSchBohBru10} and here, it does not satisfy the triangle inequality.}, $d_{\ell}$, allows to determine the number of $(\kp,0)$-limited-magnitude errors that a code could correct.
\begin{definition}[\cite{CasSchBohBru10}] \label{def:distance-km=0}
For $\vx,\vy\in \Z^n$, define $N(\vx,\vy)\eqdef\abs{\set{i ; \vx[i]>\vy[i]}}$. The distance $d_\ell$ between $\vx$ and $\vy$ is defined as
\begin{equation*}
\begin{split}
  d_\ell(\vx,\vy) \eqdef & \begin{cases}
    n+1, & \text{if $\max_i \set*{\abs{\vx[i]-\vy[i]} } > \ell$,} \\
    \max\set{N(\vx,\vy), N(\vy,\vx)} , & \text{otherwise.}
\end{cases}
\end{split}
\end{equation*}
\end{definition}

\begin{proposition}[\cite{CasSchBohBru10}]\label{prop:distance-asyerror}
A code $\cC \in \Z^n$ can correct $e$  $(\kp,0)$-limited-magnitude errors if and only if $d_{\kp}(\vx,\vy)\geq e+1$ for  all distinct $\vx,\vy\in \cC$.
\end{proposition}

If the number of errors, $t$, exceeds the unique-decoding radius, $e$, of $\cC$, the error balls of radius $t$ centered at the codewords might intersect.
For any two distinct vectors $\vx,\vy \in \Z^n$, let $N(\vx,\vy;t,\kp,\km)$ be the size of the intersection of the two balls $\vx+\BALL$ and $\vy+\BALL$, i.e.,
\[ N(\vx,\vy;t,\kp,\km) \eqdef \abs*{(\vx+\BALL)\cap(\vy+\BALL)}. \]
Given a code $\cC \subseteq \Z^n$, let $N(\cC;t,\kp,\km)$ be the size of the maximum intersection of any two balls  centered at different codewords of $\cC$, that is,
\[N(\cC;t,\kp,\km)\eqdef \max_{\substack{\vx,\vy\in \cC  \\ \vx \neq \vy }} \set*{ N(\vx,\vy {;} t,\kp,\km)}   = \max_{\substack{\vx,\vy\in \cC  \\ \vx \neq \vy }} \set*{\abs*{ \parenv{\vx+\BALL} \cap  \parenv{\vy+\BALL} }}.\]

In Section~\ref{sec:intersection} we shall first extend the definition of $d_\ell$ to capture the error-correcting capability for $\km>0$, then give some upper bounds and lower bounds on $N(\cC;t,\kp,\km)$ for any code $\cC$ of distance at least a prescribed value. In this way, we show that for any fixed integers $t$ and $e$ with $t>e$, a code $\cC$ can correct $e$ $(\kp,\km)$-limited-magnitude errors if and only  if the maximum intersection $N(\cC;t,\kp,\km)$   has  size $O(n^{t-e-1})$.

\subsection{Lattice Code/Packing and  Group Splitting}
Let $G$ be a finite Abelian group, where $+$ denotes the group
  operation. For $m\in\Z$ and $g\in G$, let $mg$ denote $g+g+\dots+g$
  (with $m$ copies of $g$) when $m>0$, which is extended in the
  natural way to $m\leq 0$.

A  \emph{lattice} is an additive subgroup $\Lambda$ of $\Z^n$. Throughout the paper we shall assume lattices are non-degenerate, namely, the quotient group $\Z^n/\Lambda$ is a finite group. The \emph{density} of $\Lambda$ is defined as $\abs{\Z^n / \Lambda }^{-1}$.  Let $G$ be an Abelian group and  $\vs=(s_1,s_2,\ldots,s_n)$ be a sequence of $G^n$. Define  \[\Lambda \eqdef \set*{\vx \in \Z^n;  \vx \cdot \vs =0}.\] Then $\Lambda$ is a lattice.
Conversely, for any lattice $\Lambda \subseteq \Z^n$, let $G$ be the quotient group, i.e., $G\eqdef \Z^n /
  \Lambda$. Let $\phi:\Z^n\to G$ be the natural homomorphism, namely
  the one that maps any $\vx\in\Z^n$ to the coset of $\Lambda$ in
  which it resides. Let $\ve_i$
  be the $i$-th unit vector in $\Z^n$ and set $s_i\eqdef\phi(\ve_i)$ for
  all $1\leq i\leq n$ and $\vs\eqdef (s_1,s_2,\dots,s_n)$. Then $\Lambda=\ker\phi=\set*{\vx \in \Z^n;  \vx \cdot \vs =0}$.

A lattice code correcting $t$ $(\kp,\km)$-limited magnitude errors is equivalent to a lattice packing of $\Z^n$ with $\BALL$. Their connection to group splitting  for $t=1$ has been observed in \cite{Sch12}.  For an excellent
treatment and history, the reader is referred to~\cite{SteSza94} and the many references therein. Recently, an extended definition of group splitting was proposed  \cite{WeiSch21} in connection with  lattice packings of  $\cB(n,t,\kp,\km)$ with $t>1$, i.e., lattice codes that correct multiple errors.

\begin{definition}[\cite{WeiSch21}]
  \label{def:split}
  Let $G$ be a finite Abelian group. Let $M\subseteq\Z\setminus\set*{0}$ be a
  finite set, and $\vs=(s_1,s_2,\dots,s_n)\in G^n$.
    If the elements $\ve\cdot \vs$, where $\ve\in
    (M\cup\set{0})^n$ and $1\leq \wt(\ve)\leq t$, are all distinct and
    non-zero in $G$, we say the
    set $M$ partially $t$-splits $G$ with a splitter vector $\vs$, denoted
    \[G \geq M\splt_t \vs.\]
\end{definition}

In our context of $(\kp,\km)$-limited-magnitude errors,  we need to take $M\eqdef
[-\km,\kp]^*$. The following  theorem shows the equivalence
of partial $t$-splittings with $M$ and lattice packings of $\BALL$.

\begin{theorem}[\cite{WeiSch21}]\label{thm:splitting}
  Let $\Lambda\subseteq\Z^n$ be a lattice. Define $G\eqdef \Z^n /
  \Lambda$. Set  $\vs\eqdef (\phi(\ve_1), \phi(\ve_2), \ldots, \phi(\ve_n)  )$,  where  $\phi:\Z^n\to G$ is the natural homomorphism.
Then $\cB(n,t,\kp,\km)$ lattice packs $\Z^n$ by $\Lambda$ if and only if  $G\geq [-\km,\kp]^*\splt_t \vs$.
\end{theorem}

\section{Maximum Intersection of Two Error Balls}\label{sec:intersection}
In this section we study the size of the maximum intersection $N(\cC;t,\kp,\km)$ for any given code $\cC \subseteq \Z^n$. This is an essential component in analyzing reconstruction codes since it determines the number of distinct channel outputs needed for the reconstruction to be successful. We first look at the case of $\cC=\Z^n$.

\begin{theorem}\label{thm:intersection-wholespace}
For any  $n,t,\kp,\km$ with $t\leq n$ and $0\leq \km\leq \kp$, we have that
\[N(\Z^n;t,\kp,\km) = \sum_{i=0}^{t-1} \binom{n-1}{i}(\kp+\km)^{i+1}=(\kp+\km)V_{\kp+\km+1}(n-1,t-1). \]
\end{theorem}

\begin{IEEEproof}
Consider the two words $\vx=(0,0,0,\ldots,0)$ and $\vy=(1,0,0,\ldots,0)$ in $\Z^n$. Then it is easy to see that the intersection of the two balls centered at $\vx$ and $\vy$ has size $(\kp+\km) \sum_{i=0}^{t-1}\binom{n-1}{i}(\kp+\km)^i$.

Denote  $N\triangleq  (\kp+\km) \sum_{i=0}^{t-1}\binom{n-1}{i}(\kp+\km)^i$. In the following, we shall show that $N(\Z^n;t,\kp,\km)\leq N$  by giving  a decoding algorithm based on the majority rule. Fix an arbitrary vector $\vx\in \Z^n$. Suppose that  $\vz_1,\vz_2,\ldots,\vz_{N+1} \in \vx+\BALL$. For each $i \in [1,n]$, define the \emph{multiset}
\[\cZ_i\eqdef \set{\vz_1[i], \vz_2[i],\ldots,\vz_{N+1}[i]}.\]
Let $m_i$ be the smallest element of $\cZ_i$ and $M_i$ be the largest element of $\cZ_i$.
If $M_i-m_i=\kp+\km$, necessarily  $\vx[i]=m_i+\km$. Otherwise, $M_i - m_i<\kp+\km$ and there are at most $\kp+\km$ distinct elements in $\cZ_i$. For each $m_i\leq  a\leq M_i$, if $a\not = \vx[i]$, then the number of $\vz_j$'s such that $\vz_j[i]=a$ is at most $\sum_{i=0}^{t-1}\binom{n-1}{i}(\kp+\km)^i$. Since $N+1=(\kp+\km)\sum_{i=0}^{t-1}\binom{n-1}{i}(\kp+\km)^{i}+1$, $\vx[i]$ must be the most frequently occurring element of $\cZ_i$.
\end{IEEEproof}

We now move on to study the case where $\cC\subseteq \Z^n$ is a code of distance at least a prescribed value. We first consider the case of $\km=0$.

\begin{lemma}\label{lm:intersection-km=0} Let $\delta \leq t\leq n$.
For any two vectors $\vx,\vy\in\Z^n$ with $d_{\kp}(\vx,\vy)=\delta \leq n$, we have that
\begin{equation}\label{eq:intersectionsize-1}
\sum_{i=0}^{t-\delta}\binom{n-2\delta}{i}  (\kp)^{i} \leq N(\vx,\vy;t,\kp,0) \leq \sum_{i=0}^{t-\delta}\binom{n-\delta}{i}  (\kp)^{i} \sum_{k=\delta+i-t}^{t-i} \binom{\delta}{k}(\kp-1)^{\delta-k}.
\end{equation}
\end{lemma}

\begin{IEEEproof}
Assume that $N(\vx,\vy)=\delta$ and $N(\vy,\vx)=\delta'$ with $\delta' \leq \delta$, where $N(\vx,\vy)\eqdef\abs{\set{i ; \vx[i]>\vy[i]}}$. Let $\vz$ be an element in $\cB_t(\vx)\cap\cB_t(\vy)$. Denote  by $i$ the number of positions where $\vx$ and $\vy$ coincide
but differ from $\vz$, denote by $j$ the number of positions where
$\vx$ and $\vz$ coincide but differ from $\vy$, and denote by $k$  the number
of positions where $\vy$ and $\vz$ coincide but differ from $\vx$. Then $0 \leq i \leq n-\delta-\delta'$, $0\leq j \leq \delta$, $0\leq k \leq \delta'$, $0\leq \delta-j+\delta'+i\leq t$, and $0\leq \delta+\delta'-k + i\leq t$. These lead to $0\leq i \leq t-\delta$, $\delta+\delta'+i-t \leq j \leq t-i$, and $\delta+\delta'+i-t \leq k \leq t-i$. Hence, the number of  choices for $\vz$ is at least
\[
\sum_{i=0}^{t-\delta}\binom{n-\delta-\delta'}{i}  (\kp)^{i};
\]
and at most
\begin{equation}\label{eq:intersectionsize}
\sum_{i=0}^{t-\delta}\binom{n-\delta-\delta'}{i}  (\kp)^{i} \sum_{j=\delta+\delta'+i-t}^{t-i} \binom{\delta}{j}  \sum_{k=\delta+\delta'+i-t}^{t-i} \binom{\delta'}{k}   (\kp-1)^{\delta+\delta'-j-k}.
\end{equation}
Thus, we just proved the lower bound, since $\binom{n-\delta-\delta'}{i} \geq \binom{n-2\delta}{i}$.
In the following, we show that \eqref{eq:intersectionsize} is decreasing with $\delta'$. Note that \eqref{eq:intersectionsize} is achieved only when  $\abs{\vx[i]-\vy[i]}\leq 1$ for all $1\leq i \leq n$. W.l.o.g., we assume that
\[\vx=(\underbrace{1,1,\ldots,1}_{\delta}, \underbrace{0,0,\ldots,0}_{\delta'}, \underbrace{0,0,\ldots,0}_{n-\delta-\delta'}),\]
and
\[\vy=(\underbrace{0,0,\ldots,0}_{\delta}, \underbrace{1,1,\ldots,1}_{\delta'}, \underbrace{0,0,\ldots,0}_{n-\delta-\delta'}).\]
Let
\[\vy'=(\underbrace{0,0,\ldots,0}_{\delta}, \underbrace{1,1,\ldots,1}_{\delta'-1}, \underbrace{0,0,\ldots,0}_{n-\delta-\delta'+1}).\]
We are going to show that
\begin{equation}\label{eq:interdecrease}
\abs*{\cB_t(\vx)\cap \cB_t(\vy)} \leq \abs*{\cB_t(\vx)\cap \cB_t(\vy')}.
\end{equation}
 For any $\vz \in (\cB_t(\vx)\cap \cB_t(\vy))\setminus (\cB_t(\vx)\cap \cB_t(\vy'))$,  we have that $\vz[\delta+\delta']\in [1,\kp]$. Furthermore, since $\vz \in \cB_t(\vy)$ and $\vz \not \in \cB_t(\vy')$, necessarily $\vz[\delta+\delta']=1$.

 Let $\vz'$ be the vector obtained from $\vz$ by changing $\vz[\delta+\delta']$ from `$1$' to `$0$'. Then it is easy to verify that $\vz'\in \cB_t(\vx)$, $\vz'\in \cB_t(\vy')$ and $\vz'\not\in \cB_t(\vy)$. Hence,
 \[\vz' \in (\cB_t(\vx) \cap \cB_t(\vy')) \setminus  (\cB_t(\vx) \cap \cB_t(\vy)).\]
 Note that for different choices of $\vz$, $\vz'$ are pairwise distinct. Therefore, we have proved \eqref{eq:interdecrease}, and so \eqref{eq:intersectionsize} is decreasing with $\delta'$. The upper bound in  \eqref{eq:intersectionsize-1} follows from \eqref{eq:intersectionsize} by taking $\delta'=0$.
\end{IEEEproof}

\begin{remark}
The lower bound in \eqref{eq:intersectionsize-1} can be attained if $N(\vx,\vy)=N(\vy,\vx)=\delta$ and ${\vx[i]-\vy[i]} \in \set{\kp,0,-\kp}$ for all $1\leq i\leq n$; the upper bound in  \eqref{eq:intersectionsize-1} can be attained if $N(\vy,\vx)=0$ and ${\vx[i]-\vy[i]} \in \set{0,1}$ for all $1\leq i\leq n$.
\end{remark}

With Lemma~\ref{lm:intersection-km=0} in hand, we can bound the intersection of balls around codewords in a general code, and with $\km=0$.

\begin{lemma}\label{lm:intersection-bound-asy}
Let $\delta\leq t \leq n$, and $\cC\subseteq \Z^n$ be a code with minimum distance $d_{\kp}(\cC)=\delta$.  Then
\[ \sum_{i=0}^{t-\delta}\binom{n-2\delta}{i}  (\kp)^{i} \leq  N(\cC;t,\kp,0) \leq \sum_{i=0}^{t-\delta}\binom{n-\delta}{i}  (\kp)^{i} \sum_{k=\delta+i-t}^{t-i} \binom{\delta}{k}(\kp-1)^{\delta-k}. \]
\end{lemma}

\begin{IEEEproof} It suffices to show that the upper bound in \eqref{eq:intersectionsize-1} is decreasing with $\delta$. The proof is the same as that  in the proof of Lemma~\ref{lm:intersection-km=0}. According to the remark after Lemma~\ref{lm:intersection-km=0}, w.l.o.g, we assume that
\[\vx=(\underbrace{1,1,\ldots,1}_{\delta},  \underbrace{0,0,\ldots,0}_{n-\delta}),\]
and
\[\vy=(\underbrace{0,0,\ldots,0}_{\delta},  \underbrace{0,0,\ldots,0}_{n-\delta}).\]
Let
\[\vx'=(\underbrace{1,1,\ldots,1}_{\delta-1},\underbrace{0,0,\ldots,0}_{n-\delta+1}).\]
For any $\vz \in (\cB_t(\vx) \cap \cB_t(\vy)) \setminus (\cB_t(\vx') \cap \cB_t(\vy))$, we have that $\vz[\delta]=1$. Let  $\vz'$ be the vector obtained from $\vz$ by changing $\vz[\delta+\delta']$ from `$1$' to `$0$'. Then $\vz' \in (\cB_t(\vx') \cap \cB_t(\vy)) \setminus (\cB_t(\vx) \cap \cB_t(\vy))$. Hence, $\abs{\cB_t(\vx) \cap \cB_t(\vy)} \leq \abs{\cB_t(\vx') \cap \cB_t(\vy)}$.
\end{IEEEproof}

\begin{theorem}
Let $\delta,t$ and $\kp$ be fixed integers such that $1\leq \delta\leq t \leq n$. For any code $\cC\subseteq \Z^n$, $N(\cC;t,\kp,0)=O(n^{t-\delta})$ if and only of $\cC$ can correct $\delta-1$ $(\kp,0)$-limited-magnitude errors.
\end{theorem}

\begin{IEEEproof} According to Theorem~\ref{thm:intersection-wholespace} and Lemma~\ref{lm:intersection-bound-asy}, $N(\cC;t,\kp,0)=O(n^{t-\delta})$ if and only if $d_{\kp}(\cC) \geq \delta$. Combining this with
 Proposition~\ref{prop:distance-asyerror}, the theorem is proved.
\end{IEEEproof}

Now, we direct our attention to the case of $\km >0$.  We first  extend  the distance $d_{\kp}$ from Definition~\ref{def:distance-km=0}.

\begin{definition}\label{def:asymdistance}
For $\vx,\vy\in \Z^n$, we define  
\begin{align*}
N_{\km}(\vx,\vy)  &\eqdef\abs*{\set*{i; 0<\abs*{\vx[i]-\vy[i]} \leq \km}}, \\ N_{\kp,\km}(\vx,\vy) &\eqdef\abs*{\set*{i; \kp<\abs*{\vx[i]-\vy[i]} \leq \kp+\km}},\\ 
M_{\kp,\km}(\vx,\vy)&\eqdef\abs*{\set*{i; \km< \vx[i]-\vy[i]\leq \kp } },
\end{align*}
where we note that $M_{\kp,\km}(\vx,\vy)$ and $M_{\kp,\km}(\vy,\vx)$ are not necessarily the same.

If  $\max_i \{\abs{\vx[i]-\vy[i]} \} > \kp+\km$, define the distance  $d_{\kp,\km}$ between $\vx$ and $\vy$ to be $n+1$; otherwise, the distance $d_{\kp,\km}$ is defined as
\begin{align*}
d_{\kp,\km}(\vx,\vy) &\eqdef\ceil*{ \frac{1}{2}\max\parenv*{  N_{\km}(\vx,\vy) -\abs*{M_{\kp,\km}(\vx,\vy)- M_{\kp,\km}(\vy,\vx)},0 }}\\
&\quad+\max\parenv*{M_{\kp,\km}(\vx,\vy), M_{\kp,\km}(\vy,\vx)}+ N_{\kp,\km}(\vx,\vy).
\end{align*}
\end{definition}
It is worth noting that when $\km=0$, the distance $d_{\kp,0}$ defined above coincides with the distance $d_{\kp}$ in Definition~\ref{def:distance-km=0}.

\begin{proposition}\label{prop:distance-error}
A code $\cC \subseteq \Z^n$ can correct $t$ $(\kp,\km)$-limited-magnitude errors if and only if $d_{\kp,\km}(\vx,\vy)\geq t+1$ for all  distinct $\vx,\vy\in \cC$.
\end{proposition}

\begin{IEEEproof}
$(\Leftarrow)$ Let $\vx,\vy \in \cC$ be two codewords, and let $\ve,\ve'\in\BALL$ be two error vectors, such that $\vx\neq\vy$ or $\ve\neq\ve'$. Assume to the contrary that $\vx+\ve=\vy+\ve'$. Then
$\vx-\vy=\ve'-\ve$, and so $d_{\kp,\km}(\vx,\vy) = d_{\kp,\km}(\ve,\ve')$. Let  $n_1=N_{\km}(\ve,\ve')$, $n_2=N_{\kp,\km}(\ve,\ve')$, $m_1=M_{\kp,\km}(\ve,\ve')$, and $m_2=M_{\kp,\km}(\ve',\ve)$.  W.l.o.g., assume that $m_1\geq m_2$.  Since both $\ve,\ve'$ have Hamming weight at most $t$ and $\ve,\ve'\in [-\km,\kp]^n$, then 
\[m_1+n_2\leq t.\]
Next, partition the set $\cN_1\eqdef \set*{i; 0<\abs*{\vx[i]-\vy[i]} \leq \km}$ into $P_1\eqdef\set*{i;\ve'[i]<\ve[i]} \cap \cN_1$ and $P_2\eqdef\set{i;\ve'[i]>\ve[i]}\cap \cN_1$. Then we have $\abs{P_1}+m_1+n_2\leq t$ and $\abs{P_2}+m_2+n_2\leq t$.  Thus,
\[n_1+(m_1+n_2)+(m_2+n_2)  \leq 2t.\]
Hence, $d_{\kp,\km}(\ve,\ve')=\ceil{\max\set{n_1-m_1+m_2,0}/2}+m_1+n_2\leq t$, which contradicts that $d_{\kp,\km}(\vx,\vy)\geq t+1$.

$(\Rightarrow)$ Suppose that  there are two distinct codewords $\vx,\vy \in \cC$ such that $d_{\kp,\km}(\vx,\vy)\leq t$. Since $d_{\kp,\km}$ is symmetric, assume w.l.o.g.~that $M_{\kp,\km}(\vx,\vy)\leq M_{\kp,\km}(\vy,\vx)$.
Denote,
\begin{align*}
\cN_1&\eqdef\set*{i; 0<\abs{\vx[i]-\vy[i]}\leq \km  }, \\
\cN_2&\eqdef\set*{i; \kp<\abs{\vx[i]-\vy[i]}\leq \kp+\km }, \\
\cM_1&\eqdef\set*{i; \km< \vx[i]-\vy[i] \leq \kp }, \\
\cM_2&\eqdef\set*{i; \km< \vy[i]-\vx[i] \leq \kp }.
\end{align*}
Take an arbitrary subset  $\cN_1'\subseteq \cN_1$ of size $\ceil*{ \frac{1}{2}\max\set*{  N_{\km}(\vx,\vy) +M_{\kp,\km}(\vx,\vy)- M_{\kp,\km}(\vy,\vx),0 }}$. Let $\ve$ be the vector with support set $\cN_1'\cup \cN_2 \cup \cM_2$, where
$\ve[i]=\vy[i]-\vx[i]$ when $i\in \cN_1' \cup \cM_2$,  $\ve[i]=\vy[i]-\vx[i] -\km$ when $\vy[i]-\vx[i]>\kp$ and $\ve[i]=\vy[i]-\vx[i] +\kp$ when $\vy[i]-\vx[i]<-\kp$.
Let $\ve'$ be the vector with support set $(\cN_1\setminus \cN_1')\cup \cN_2 \cup \cM_1$, where
$\ve'[i]=\vx[i]-\vy[i]$ when $i\in (\cN_1\setminus \cN_1')\cup \cM_1$,  $\ve'[i]=-\km$ when $\vy[i]-\vx[i]>\kp$ and $\ve'[i]=\kp$ when $\vy[i]-\vx[i]<-\kp$.

It follows that $\ve'=\vx-\vy+\ve$, i.e., $\vx+\ve=\vy+\ve'$, and  $\ve,\ve'\in[-\km,\kp]^n$.
In the following, we verify that both $\ve$ and $\ve'$ have Hamming weight at most $t$, which contradicts the fact that $\cC$ can correct $t$ $(\kp,\km)$-limited-magnitude errors.
The vector $\ve$ has Hamming weight
 \begin{align*}
 \abs{\cN_1'\cup \cM_2 \cup \cN_2} & =\ceil*{ \frac{1}{2}\max\set*{  N_{\km}(\vx,\vy) +M_{\kp,\km}(\vx,\vy)- M_{\kp,\km}(\vy,\vx),0 }}+ M_{\kp,\km}(\vy,\vx)+ N_{\kp,\km}(\vx,\vy)\\
 &=d_{\kp,\km}(\vx,\vy)\leq t.
 \end{align*}
 For the vector $\ve'$, if $N_{\km}(\vx,\vy) +M_{\kp,\km}(\vx,\vy)- M_{\kp,\km}(\vy,\vx)\leq 0$, then it has Hamming weight
 \begin{align*}
 \abs{ (\cN_1\setminus \cN_1')\cup \cM_1 \cup \cN_2  } &=N_{\km}(\vx,\vy)+M_{\kp,\km}(\vx,\vy)+N_{\kp,\km}(\vx,\vy) \leq M_{\kp,\km}(\vy,\vx)+N_{\kp,\km}(\vx,\vy) \\
 &=d_{\kp,\km}(\vx,\vy)\leq t.
 \end{align*}
Otherwise, it has Hamming weight
\begin{align*}
 \abs{ (\cN_1\setminus \cN_1')\cup \cM_1 \cup \cN_2  } &\leq \frac{1}{2}\parenv*{ N_{\km}(\vx,\vy) -M_{\kp,\km}(\vx,\vy)+ M_{\kp,\km}(\vy,\vx)}+M_{\kp,\km}(\vx,\vy)+N_{\kp,\km}(\vx,\vy)  \\
  &= \frac{1}{2}\parenv*{ N_{\km}(\vx,\vy) +M_{\kp,\km}(\vx,\vy)- M_{\kp,\km}(\vy,\vx)}+M_{\kp,\km}(\vy,\vx)+N_{\kp,\km}(\vx,\vy)  \\
 &\leq d_{\kp,\km}(\vx,\vy)\leq t.
 \end{align*}
\end{IEEEproof}

\begin{lemma}\label{lm:intersection-ubound}
 Assume that $\delta \leq t\leq n$ and $0<\km\leq \kp$. For any two vectors $\vx,\vy\in \Z^n$ with $d_{\kp,\km}(\vx,\vy)=\delta$, we have that
\begin{equation}\label{eq:intersection-ubound}
\sum_{i=0}^{t-\delta}\binom{n-2\delta}{i} (\kp+\km)^{i}\leq N(\vx,\vy;t,\kp,\km)\leq \sum_{i=0}^{t-\delta}\binom{n}{i} (\kp+\km)^{i+2\delta}.
\end{equation}
\end{lemma}

\begin{IEEEproof} Let $n_1=N_{\km}(\vx,\vy)$, $n_2=N_{\kp,\km}(\vx,\vy)$, $m_1 =M_{\kp,\km}(\vx,\vy)$, and $m_2 =M_{\kp,\km}(\vy,\vx)$. The maximal intersection is achieved if $\vx$ and $\vy$ have the following form:
\[\vx = ( \underbrace{0,0,\ldots,0}_{n_1},\underbrace{0,0,\ldots,0}_{n_2},\underbrace{\km+1,\km+1,\cdots,\km+1}_{m_1}, \underbrace{0,0,\ldots,0}_{m_2}, \underbrace{0,0,\ldots,0}_{n'} )\]
and
\[\vy = ( \underbrace{1,1,\ldots,1}_{n_1},\underbrace{\kp+1,\ldots,\kp+1}_{n_2},\underbrace{0,0,\ldots,0}_{m_1},\underbrace{\km+1,\cdots,\km+1}_{m_2},  \underbrace{0,0,\ldots,0}_{n'}).\]
 where $n'=n-n_1-n_2-m_1-m_2$.
Partition the positions into the following five intervals:
\begin{align*}
I_1=[1,n_1], I_2=[n_1+1,n_1+n_2], I_3=[n_1+n_2+1,n_1+n_2+m_1], \\
 I_4=[n_1+n_2+m_1+1,n_1+n_2+m_1+m_2], I_5=[n-n'+1,n].
 \end{align*}

 Let $\vz$ be an element in the intersection of the two balls $\cB_t(\vx)$ and $\cB_t(\vy)$.
 Obviously, $\vx$ and $\vz$ can have the same components only in the positions belonging to $I_1\cup I_3 \cup I_5$, and $\vy$ and $\vz$ can have the same components only in the positions belonging to $I_1\cup I_4 \cup I_5$.
 Denote  by $i$ the number of positions where $\vx$ and $\vy$ coincide
but differ from $\vz$; so all these $i$ positions come from $I_5$. Denote
\begin{align*}
 j & = \abs*{\set*{\ell \in I_1; \vz[\ell] =\vx[\ell] }}, \\
 k & = \abs*{\set*{\ell \in I_1; \vz[\ell] =\vy[\ell] }}, \\
 r & = \abs*{\set*{\ell \in I_3; \vz[\ell] =\vx[\ell] }}, \\
 s & = \abs*{\set*{\ell \in I_4; \vz[\ell] =\vy[\ell] }}.
\end{align*}
Then we have $j+k\leq n_1$, $r\leq m_1$, $s\leq m_2$, $n_1-j+n_2+m_1-r+m_2+i \leq t$, and $n_1-k+n_2+m_1+m_2-s+i \leq t$. Hence,
\begin{align*}
i & \leq  \min\set*{ t-\ceil*{\frac{n_1+m_1+m_2}{2}}-n_2, t-m_1-n_2,t-m_2-n_2 }\\
&= t-\max\set*{ \ceil*{\frac{n_1+m_1+m_2}{2}}+n_2, m_1+n_2,m_2+n_2 }\\
&=t- d_{\kp,\km}(\vx,\vy).
\end{align*}
Thus, the number of choices of $\vz$ is at most
\begin{align*}
&\sum_{i=0}^{t-\delta} \binom{n'}{i}(\kp+\km)^i  \sum_j\binom{n_1}{j} \sum_k  \binom{n_1-j}{k} (\kp+\km-2)^{n_1-j-k} (\km)^{n_2} \\
& \times\sum_r \binom{m_1}{r} \sum_s \binom{m_2}{s}(\kp)^{m_1+m_2-r-s}   \\
\leq &  \sum_{i=0}^{t-\delta} \binom{n'}{i}   (\kp+\km)^{i+n_1} (\km)^{n_2} (\kp+1)^{m_1+m_2}\\
 < & \sum_{i=0}^{t-\delta} \binom{n}{i}   (\kp+\km)^{i+2\delta} . \end{align*}

 For the lower bound, we have that
 \[N(\vx,\vy;t,\kp,\km) \geq \sum_{i=0}^{t-\delta} \binom{n'}{i}(\kp+\km)^i \geq \sum_{i=0}^{t-\delta} \binom{n-2\delta}{i}(\kp+\km)^i. \]
\end{IEEEproof}

\begin{theorem}\label{thm:intersection-ercrtcap}
  Let $\kp,\km,t$ and $\delta$ be fixed integers such that  $0<\km\leq \kp$ and $1\leq \delta \leq t$.
For any code $\cC \subseteq \Z^n$,  $N(\cC;t,\kp,\km)=O(n^{t-\delta})$ if and only if $\cC$ can correct  $\delta-1$ $(\kp,\km)$-limited-magnitude errors.
\end{theorem}

\begin{IEEEproof}
According to Lemma~\ref{lm:intersection-ubound}, $N(\cC;t,\kp,\km)=O(n^{t-\delta})$ if and only if $d_{\kp,\km}(\cC)\geq\delta$. Combining this with Proposition~\ref{prop:distance-error}, the theorem is proved.
\end{IEEEproof}

\section{Single-Error Lattice Reconstruction Codes}\label{sec:codedesign}

In this section, we study the design of reconstruction codes from a given number received sequences. In other words, given a positive integer $N$, we would like to construct a code $\cC \subseteq \Z^n$ such that $N(\cC;t,\kp,\km)\leq N$. Since the general problem seems to be involved, we focus on the case of lattice codes in the channel which introduces a single $(\kp,\km)$-limited-magnitude error. Theorem~\ref{thm:intersection-wholespace} shows that if $N\geq \kp+\km$, we can take the whole space $\Z^n$ as our code, and  its density is $1$. In the other extremal case of $N=0$, Theorem~\ref{thm:splitting} and Definition~\ref{def:split} imply  that any lattice code correcting a single   error should have density at most $(n(\kp+\km)+1)^{-1}$, which is  $O(1/n)$.

In the following, we study the case of $N=1$, and since lattice codes are in question, we base our approach on group splitting.

\begin{lemma} Assume that $1\leq \km\leq \kp$. Let $\Lambda\subseteq\Z^n$ be a lattice. Define $G= \Z^n /
  \Lambda$, and  $s_i =\phi(\ve_i)$ for $1\leq i \leq n$,  where  $\phi:\Z^n\to G$ is the natural homomorphism. Let $\vs=(s_1,s_2,\ldots,s_n)$ and so $\Lambda=\set{\vx\in \Z^n; \vx\cdot \vs=0}$.
Then    $N(\Lambda;1,\kp,\km)\leq 1$ if and only if all the following hold:
\begin{enumerate}
 \item[(C1)] $a s_i \not =0$ for all $1\leq i \leq n$ and $a  \in [-\km,\kp]^*$.
 \item[(C2)] $a s_i \not = b s_i$ for all $1\leq i \leq n$ and  all distinct $a,b\in [-\km,\kp]^*$, except $\abs{a-b}=\kp+\km$.
 \item[(C3)] $a s_i \not = b s_j$ for all $1\leq i <j \leq n$ and all $a,b\in [-\km,\km]^*$.
\end{enumerate}
\end{lemma}

\begin{IEEEproof}
We first show that if $N(\Lambda;1,\kp,\km)\leq 1$, then  the  conditions hold.
\begin{enumerate}
  \item For (C1), w.l.o.g., suppose to the contrary that $a s_1  = 0$  for some  $a\in [-\km,\kp]^*$.   Then both the vectors $\vx=(0,0,\ldots,0)$ and $\vy=(a,0,\ldots,0)$  belong to $\Lambda$. If $a\leq \km$, then $\set{\vx,\vy} \subseteq \cB_1(\vx)\cap \cB_1(\vy)$; otherwise, the intersection contains $(a,0,\ldots,0)$ and $(a-1,0,\ldots,0)$.
  \item For (C2), w.l.o.g., suppose to the contrary that $a s_1  = b s_1$  for some  $a,b\in [-\km,\km]^*$ with $b>a$ and $\abs{a-b}<\kp+\km$. Consider the two codewords $\vx=(0,0,\ldots,0)$ and $\vy=(b-a,0,\ldots,0)$. Since $1\leq b-a \leq \kp+\km-1$, then $b-a-\km,b-a-\km+1 \in [-\km,\kp]$. Hence, the intersection $ \cB_1(\vx)\cap \cB_1(\vy)$ contains two vectors $(b-a-\km,0,\ldots,0)$ and  $(b-a-\km+1,0,\ldots,0)$, a contradiction.
   \item For (C3), suppose to the contrary that $a s_i  = b s_j$  for some $1\leq i <j \leq n$ and $a,b\in [-\km,\km]^*$. W.l.o.g., we assume that $i=1$ and $j=2$. Then both the vectors $\vx=(0,0,\ldots,0)$ and $\vy=(a,-b,0,\ldots,0)$  belong to $\Lambda$ as $\vx\cdot \vs = \vy \cdot \vs=0$. However, the intersection $\cB_1(\vx)\cap \cB_1(\vy)$ contains two vectors  $(a,0,0,\ldots,0)$ and $(0,-b,0,\ldots,0)$, which contradicts $N(\Lambda;1,\kp,\km)\leq1$.
\end{enumerate}

Now, we show the other direction.  Assume that $\vx,\vy\in \Lambda$, $\vx\neq\vy$, with $ \cB_1(\vx)\cap \cB_1(\vy)\neq \varnothing$. Then  $\vx+a\ve_i= \vy+b\ve_j$ for some $1\leq i,j\leq n$ and $a,b\in[-\km,\kp]$. Hence, $a s_i- b s_j=(a\ve_i-b\ve_j)\cdot \vs =(\vy-\vx)\cdot \vs=0$, and so $a s_i=b s_j$. We consider the following two cases.
 \begin{enumerate}
 \item If $i\neq j$, according to (C1) and (C3), necessarily $a,b\neq 0$ and $\max\set{a,b}>\km$. W.l.o.g., assume that $i=1$, $j=2$, and $b > \km$. Then if $\vx=(x_1,x_2,\ldots,x_n)$, we have $\vy=(x_1+a,x_2-b,x_3,\ldots,x_n)$. Since $b>\km$, the intersection of the two balls only contains a unique vector, i.e., $(x_1+a,x_2,\ldots,x_n)$.
 \item If $i=j$, according to (C1) and (C2), necessarily $\abs{b-a}=\kp+\km$. W.l.o.g., assume that $i=j=1$ and $b-a=\kp+\km$. Then if $\vx=(x_1,x_2,\ldots,x_n)$, we have $\vy=(x_1-(\kp+\km),x_2,\ldots,x_n)$. Thus the intersection  only contains the  vector $(x_1-\km,x_2,x_2,\ldots,x_n)$.
 \end{enumerate}
\end{IEEEproof}

\begin{corollary}
Assume that $1\leq \km\leq \kp$. Let $\Lambda\subseteq\Z^n$ be a lattice code such that $N(\Lambda;1,\kp,\km)\leq 1$. Then
\begin{equation*}
\begin{split}
  \abs{\Z^n / \Lambda} \geq & \begin{cases}
     \max\set{2n\km+1,\kp+\km}, & \text{if $\kp>\km$,} \\
    \max\set{n(\kp+\km-1)+1,\kp+\km}. & \text{ if $\kp=\km$.}
\end{cases}
\end{split}
\end{equation*}
\end{corollary}

\begin{IEEEproof}
Note that the conditions (C1)--(C3) are equivalent to the following two conditions:
\begin{enumerate}
 \item[(C1')] $a s_i \not = b s_i$ for all $1\leq i \leq n$ and  all distinct $a,b\in [-\km,\kp-1]$.
 \item[(C2')] $a s_i \not = b s_j$ for all $1\leq i <j \leq n$ and all $a,b\in [-\km,\km]$, except $a=b=0$.
\end{enumerate}
For any $\vx\in\Z^n$, we can think of $\vx\cdot\vs$ as a syndrome. Thus, the elements of $\Lambda$ are exactly those with the $0$ syndrome, and the elements of a coset $\vv+\Lambda$ are exactly those with syndrome $\vv\cdot\vs$. Hence,
\[ \abs*{\Z^n/\Lambda} \geq \abs*{\set*{as_i ; a\in[-\km,\kp], 1\leq i\leq n}}.\]
The bound now follows by (C1') and (C2').
\end{IEEEproof}

From the bound above, we can see that if $\km >0$ and $\kp,\km$ are fixed, any lattice code with $N(\Lambda;1,\kp,\km)\leq 1$ has density at most $O(1/n)$, which is asymptotically the same as the case of  $N(\Lambda;1,\kp,\km)= 0$.

For $\km=0$, however, it is much different. There are  codes  with $N(\Lambda;1,\kp,0)\leq 1$ having constant density. This is trivially true for $\kp=1$ since then the lattice $\Lambda=\Z^n$ has density $1$ and it satisfies $N(\Z^n;1,1,0)=1$. For $\kp\geq 2$ we have the following:

\begin{lemma}\label{lm:recon-km=0} Assume that  $\km=0$ and $\kp\geq 2$.
Let $\Lambda\subseteq\Z^n$ be a lattice. Let $G\eqdef \Z^n / \Lambda$ and  $s_i=\phi(\ve_i)$ for $1\leq i \leq n$,  where  $\phi:\Z^n\to G$ is the natural homomorphism.  Let $\vs=(s_1,s_2,\ldots,s_n)$ and so $\Lambda=\set{\vx\in \Z^n; \vx\cdot \vs=0}$.
Then    $N(\Lambda;1,\kp,0)\leq 1$ if and only if  $a s_i \not = b s_i$ for all $1\leq i \leq n$ and  all distinct $a,b\in [0,\kp-1]$.
\end{lemma}

\begin{IEEEproof}
 ($\Rightarrow$) Suppose to the contrary that $a s_i  = b s_i$  for some  $a,b\in [0,\kp-1]$ with $b>a$. W.l.o.g., assume $i=1$. Consider the two codewords $\vx=(0,0,\ldots,0)$ and $\vy=(b-a,0,\ldots,0)$. Since $1\leq b-a \leq \kp-1$,  the intersection $ \cB_1(\vx)\cap \cB_1(\vy)$ contains the two vectors $(b-a,0,\ldots,0)$ and  $(b-a+1,0,\ldots,0)$, a contradiction.

 ($\Leftarrow$) Assume that $\vx,\vy\in \Lambda$, $\vx\neq\vy$, with $ \cB_1(\vx)\cap \cB_1(\vy)\neq \varnothing$. Then  $\vx+a\ve_i= \vy+b\ve_j$ for some $1\leq i,j\leq n$ and $a,b\in[0,\kp]$. So $a s_i- b s_j=(a\ve_i-b\ve_j)\cdot \vs =(\vy-\vx)\cdot \vs=0$, hence, $a s_i=b s_j$. We consider the following two cases.
 \begin{enumerate}
 \item If $i\neq j$ and $a,b\neq 0$, w.l.o.g., assume that $i=1$, $j=2$ and $b \geq a$. Then if $\vx=(x_1,x_2,\ldots,x_n)$, we have $\vy=(x_1+a,x_2-b,x_3,\ldots,x_n)$.
 Since $a,b>0$,  the intersection of the two balls only contains the vector $(x_1+a,x_2,\ldots,x_n)$. If $a=0$, then $\vx+0\cdot\ve_i=\vx+0\cdot\ve_j=\vx=\vy+b\ve_{j}$, which is included in the case $i=j$, and a symmetric argument applies to the case of $b=0$.
 \item If $i=j$, according to our condition, $(a,b)=(0,\kp)$ or $(\kp,0)$. In both cases,  the  intersection  only contains one  vector, i.e., $\vx$ or $\vy$ respectively.
 \end{enumerate}
\end{IEEEproof}

\begin{corollary}
Assume that $\km=0$ and $\kp\geq 2$. Let $\Lambda\subseteq\Z^n$ be a lattice code such that $N(\Lambda;1,\kp,0)\leq 1$. Then
\[\abs{\Z^n/\Lambda} \geq \kp.\]
Moreover, the bound can be attained by letting $G=\Z_{\kp}$ and $\vs=(1,1,\ldots,1)$, where the corresponding lattice code is
\[ \Lambda=\set*{(x_1,x_2,\ldots,x_n)\in \Z^n; \sum_{i=1}^n x_i\equiv 0 \pmod{\kp}}.\]
\end{corollary}

\begin{IEEEproof} The bound comes directly from Lemma~\ref{lm:recon-km=0}.
For the code $\Lambda$, since $s_i=1$,  $as_i\not \equiv 0 \pmod{\kp}$ for all $a  \in [1,\kp-1]$. Thus, according to Lemma~\ref{lm:recon-km=0}, we have $N(\Lambda;1,\kp,0)\leq 1$.
\end{IEEEproof}

Now,  we study the case of $N(\Lambda;1,\kp,\km)\leq 2$ with $\km\geq 1$. We have the following result which is similar to the case of  $N(\Lambda;1,\kp,0)\leq 1$. This time, $(\kp,\km)=(1,1)$ is a trivial case in which we can take $\Lambda=\Z^n$ since $N(\Z^n;1,1,1)=2$. The non-trivial cases are given by the following:

\begin{lemma}  Assume that $1\leq \km\leq \kp$ and $\kp+\km\geq 3$. Let $\Lambda\subseteq\Z^n$ be a lattice. Let $G\eqdef \Z^n / \Lambda$ and  $s_i=\phi(\ve_i)$ for $1\leq i \leq n$,  where  $\phi:\Z^n\to G$ is the natural homomorphism.  Let $\vs=(s_1,s_2,\ldots,s_n)$ and so $\Lambda=\set{\vx\in \Z^n; \vx\cdot \vs=0}$.
Then    $N(\Lambda;1,\kp,\km)\leq 2$ if and only if  $a s_i \not = b s_i$ for all $1\leq i \leq n$ and  all distinct $a,b\in [-\km,\kp-2]$.
\end{lemma}

\begin{IEEEproof}
 ($\Rightarrow$) Suppose to the contrary that $a s_i  = b s_i$  for some  $a,b\in [-\km,\kp-2]$. W.l.o.g., assume that $i=1$ and $b>a$. Consider the two codewords $\vx=(0,0,\ldots,0)$ and $\vy=(b-a,0,\ldots,0)$. Since $1\leq b-a \leq \kp+\km-2$,  the intersection $ \cB_1(\vx)\cap \cB_1(\vy)$ contains three vectors $(b-a-\km,0,\ldots,0)$ and  $(b-a-\km+1,0,\ldots,0)$ and $(b-a-\km+2,0,\ldots,0)$, a contradiction.

 ($\Leftarrow$) Assume that $\vx,\vy\in \Lambda$, $\vx\neq\vy$, with $ \cB_1(\vx)\cap \cB_1(\vy)\neq \varnothing$. Then  $\vx+a\ve_i= \vy+b\ve_j$ for some $1\leq i,j\leq n$ and $a,b\in[-\km,\kp]$. Hence, $a s_i- b s_j=(a\ve_i-b\ve_j)\cdot \vs =(\vy-\vx)\cdot \vs=0$, and so $a s_i = b s_j$. We consider the following two cases:
 \begin{enumerate}
 \item If $i\neq j$ and $a,b\neq 0$, i.e., $\vx$ and $\vy$ differ in two positions, necessarily  the intersection of the two balls  contains at most two vectors. If either $a=0$ or $b=0$ then the case is covered by the following case of $i=j$.
 \item If $i=j$, according to our assumption, $\abs{a-b}=\kp+\km-1$ or $\kp+\km$. If $\abs{a-b}=\kp+\km$,  the  intersection  only contains one  vector, and if $\abs{a-b}=\kp+\km-1$,  the  intersection   contains two  vectors.
 \end{enumerate}
\end{IEEEproof}

\begin{corollary}
Assume that $1\leq \km\leq \kp$ and $\kp+\km\geq 3$.  Let $\Lambda\subseteq\Z^n$ be a lattice code such that $N(\Lambda;1,\kp,\km)\leq 2$. Then
\[\abs{\Z^n/\Lambda} \geq \kp+\km-1.\]
Moreover, the bound can be attained by letting $G=\Z_{\kp+\km-1}$ and $\vs=(1,1,\ldots,1)$, where the corresponding code is
\[ \Lambda=\set*{(x_1,x_2,\ldots,x_n)\in \Z^n; \sum_{i=1}^n x_i\equiv 0 \pmod{\kp+\km-1}}.\]
\end{corollary}

\section{Efficient Reconstruction Algorithms}\label{sec:algorithm}
In this section, we present two reconstruction algorithms for the $(\kp,\km)$-limited-magnitude errors. We assume nothing about the structure of the code. In particular, we do not assume the codes are linear, i.e., lattice codes. Since more errors may occur in the received vectors than that are correctable by unique decoding, our strategy is to combine the received vectors into a single vector that is guaranteed to be within the unique-decoding radius from the transmitted codeword, and then use a unique decoding procedure. We thus reduce the reconstruction problem to a classical decoding problem.

If $\cC\subseteq\Z^n$ is a code capable of correcting $\delta-1$ $(\kp,\km)$-limited-magnitude errors, we assume the existence of a decoding function $\cD_\cC:\Z^n\to\cC$ which upon receiving a codeword corrupted by at most $\delta-1$ $(\kp,\km)$-limited-magnitude errors, is capable of finding the transmitted codeword. If there exists an efficient test of whether a vector is in $\cC$, as in the case of lattice codes, then a naive brute-force implementation of a decoding procedure is possible in time complexity $O(\abs{\cB(n,\delta-1,\kp,\km)})=O(n^{\delta-1})$ by testing all the vectors in a ball centered at the received vector. 

The first algorithm we present only works for the case of $\km=0$ and requires a few more received vectors than the upper bound of Lemma~\ref{lm:intersection-bound-asy}. However, it is quite simple.

\begin{theorem}\label{thm:algorithm-km=0} Let $\cC \subseteq \Z^n$ be a code with the minimum distance $d_{\kp}(\cC)=\delta$.
Denote
\[N\triangleq (\kp)^\delta \sum_{i=0}^{t-\delta} \binom{n-\delta}{i}(\kp)^i+1=(\kp)^\delta V_{\kp+1}(n-\delta,t-\delta)+1.\]
Let $\vy_1,\vy_2,\ldots,\vy_{N}$ be $N$ distinct vectors that come from the same ball $\vx+\cB(n,t,\kp,0)$ for some codeword $\vx \in \cC$.
Then we can reconstruct $\vx$ from $\cY\eqdef\set{\vy_1,\dots,\vy_N}$ with time complexity $O(Nn+C)$, where $C$ is the time complexity of the unique-decoding algorithm of $\cC$.
\end{theorem}

\begin{IEEEproof} Our reconstruction algorithm is summarized in  Algorithm~\ref{alg:minimum}. 
Since $z_i=\min\{\vy_1[i],\vy_2[i],\ldots,\vy_{N}[i]\}$ for each $i \in [1,n]$, it is easy to see that $ 0\leq z_i-\vx[i] \leq \kp$. Thus, $\vz \in \vx+\cB(n,r,\kp,0)$ for some  integer $r$. Assume to the contrary that there are $\delta$ positions where all the vectors in $\cY$ differ from $\vx$ on each of these positions. However, the number of such vectors is at most $(\kp)^\delta \sum_{i=0}^{t-\delta} \binom{n-\delta}{i}(\kp)^i$, which is strictly less than $N$. Thus, we can find $r\leq \delta-1$, and $\vz \in \vx+ \cB(n,\delta-1,\kp,0)$. Since $\cC$ has distance $\delta$, we can run the unique-decoding algorithm, $\cD_{\cC}$, on $\vz$ to recover $\vx$.
\end{IEEEproof}

\begin{algorithm}[ht]
  \caption{Reconstruction algorithm for $\km=0$}
  \label{alg:minimum}
  \begin{algorithmic}[1]
    \Statex \textbf{Input:} an $N$-set $\cY=\set{\vy_1,\vy_2,\ldots,\vy_N} \subseteq {\vx} +\cB(n,t,\kp,0)$ for some $\vx\in\cC$
    \Statex \textbf{Output:} the codeword ${\vx}\in \cC$
    \For{$1\leq i\leq n$} 
    \State $z_i \gets \min\set{\vy_1[i], \vy_2[i],\ldots,\vy_N[i]}$
    \EndFor
    \State $\vz\gets (z_1,z_2,\ldots,z_n)$
    \State $\vx \gets \cD_{\cC}(\vz)$
    \State \Return{$\vx$}
  \end{algorithmic}
\end{algorithm}

\begin{remark}
The number of required vectors in Algorithm~\ref{alg:minimum} is larger than the upper bound of Lemma~\ref{lm:intersection-bound-asy} by, at most, a constant multiplicative factor, since the inner expression of $\sum_{k=\delta+i-1}^{t-i} \binom{\delta}{k} (\kp-1)^{\delta-k}$ in the upper bound of Lemma~\ref{lm:intersection-bound-asy} is replaced here with $\kp^{\delta}$.
\end{remark}

In the following, we consider the case of  $\km>0$. Our method is to modify the reconstruction algorithm from~\cite{SinYaa19} which was suggested for a channel with substitutions.

For a finite multiset $\cM$ with elements from $\Z$, denote $n_i(\cM)$  the number of times that  the element $i$  appears in $\cM$. Denote $\Maj(\cM)$ the element  which appears most frequently in $\cM$. If there is more than one such element, we take the smallest one.

Let $\cC\subseteq \Z^n$ be a code with $d_{\kp,\km}(\cC) =\delta$. Let $\vx \in \cC$ be a codeword and $\cY$ be a subset of  $\vx+\BALL$ with $N$ vectors, where
 \begin{equation}\label{eq:decnumofread}
 N\eqdef \sum_{i=0}^{t-\delta}\binom{n}{i} (\kp+\km)^{i+2\delta}+1=(\kp+\km)^{2\delta}V_{\kp+\km+1}(n,t-\delta)+1.
 \end{equation}
 Denote
 \begin{equation}\label{eq:decthreshold}
 \tau  \eqdef \parenv*{1-\frac{2}{\delta}}N + \frac{2}{\delta} \sum_{i=0}^{t-\delta} \binom{n-\delta}{i} (\kp+\km)^{i+\delta}=\parenv*{1-\frac{2}{\delta}}N + \frac{2(\kp+\km)^\delta}{\delta}V_{\kp+\km+1}(n-\delta,t-\delta).
 \end{equation}
It is easy to check that $\tau <N$.

 We apply the following majority algorithm (Algorithm~\ref{alg:majority}) with threshold $\tau$ on $\cY$ to get an estimate $\vz$ of $\vx$. The returned estimate may also contain the symbol $?$ which indicates an erasure.

\begin{algorithm}[ht]
  \caption{Majority algorithm}
  \label{alg:majority}
  \begin{algorithmic}[1]
    \Statex \textbf{Input:} an $N$-set $\cY=\set{\vy_1,\vy_2,\ldots,\vy_N} \subseteq  \Z^n$ and a threshold  $\tau$
    \Statex \textbf{Output:} a word $\vz\in (\Z\cup \set{?})^n$
    \For{$1\leq i\leq n$} 
    \State $\cY_i \gets \set{\vy_1[i], \vy_2[i],\ldots,\vy_N[i]}$ (multiset)
    \State $M_i\gets \Maj(\cY_i)$
    \If{ $2n_{M_i}(\cY_i)-N>\tau$}
    \State $\vz[i] \gets M_i$
    \Else
    \State $\vz[i] \gets ?$
    \EndIf
    \EndFor
    \State \Return{$\vz$}
  \end{algorithmic}
\end{algorithm}

 We have the following upper bounds on the number of errors and erasures in the estimate $\vz$.

\begin{lemma}\label{lm:boundonerror}
Let $\cY\subseteq \vx+\BALL$ be an $N$-set, $\vx\in\Z^n$, and let $\vz$ be the output of Algorithm~\ref{alg:majority} when run on $\cY$ with $\tau$ from~\eqref{eq:decthreshold}. Then $\vz$ contains at most $\delta-1$ errors compared with $\vx$, that is,
\[ \abs*{\set*{ i\in[1,n] ; \vx[i]\neq \vz[i], \vz[i]\in\Z}}\leq \delta-1.\]
\end{lemma}

\begin{IEEEproof}Suppose to the contrary that there are at least $\delta$ errors. W.l.o.g., we assume that the first $\delta$ symbols of $\vz$ are erroneous. Let $M\eqdef[-\km,\kp]^*$. For each $i\in[1,n]$ and $k \in M$, let
\[e_i^k =\abs*{\set*{\ell\in [1,N];\vy_{\ell}[i]=\vx[i]+k}}.\]
Then there is an error in the $i$-th position of $\vz$   only if there is a $k\in M$ such that $2e_i^{k}-N> \tau$. It follows that
\begin{align}\label{eq:sumerror-lowbound}
\sum_{i=1}^{\delta} \sum_{ k\in M} e_i^k & > \delta \frac{N+\tau}{2} = (\delta-1)N+ \sum_{i=0}^{t-\delta} \binom{n-\delta}{i} (\kp+\km)^{i+\delta}.
\end{align}

On the other hand, there are at most $\sum_{i=0}^{t-\delta} \binom{n-\delta}{i} (\kp+\km)^{i+\delta}$ vectors in $\cY$ that can have erroneous components in all of the first $\delta$ positions. For the other vectors in $\cY$, each  has at most $\delta-1$ errors in the first $\delta$ positions. Therefore,
\begin{align*}
\sum_{i=1}^{\delta} \sum_{ k\in M} e_i^k  & \leq  \delta \sum_{i=0}^{t-\delta} \binom{n-\delta}{i} (\kp+\km)^{i+\delta} + (\delta-1) \parenv*{N- \sum_{i=0}^{t-\delta} \binom{n-\delta}{i} (\kp+\km)^{i+\delta}} \\
&=  (\delta-1)N+ \sum_{i=0}^{t-\delta} \binom{n-\delta}{i} (\kp+\km)^{i+\delta},
\end{align*}
which contradicts \eqref{eq:sumerror-lowbound}.
\end{IEEEproof}

\begin{lemma}\label{lm:bounderasure}
Let $\cY\subseteq \vx+\BALL$ be an $N$-set, $\vx\in\Z^n$, and let $\vz$ be the output of Algorithm~\ref{alg:majority} when run on $\cY$ with $\tau$ from~\eqref{eq:decthreshold}. Then $\vz$ contains at most $2t\delta$ erasures.
\end{lemma}

\begin{IEEEproof}
Let $e_i^k$ be as in the proof of Lemma~\ref{lm:boundonerror}. Noting that $\vz[i]=\vx[i]$ if and only if $N-2\sum_{k\in M} e_i^k>\tau$, there is an erasure on the $i$-th positions only if $\sum_{k\in M} e_i^k\geq \frac{N-\tau}{2}$. Since each vector $\vy_\ell$ has at most $t$ errors, the number of erasures is at most
\begin{align*}
\frac{2t N}{{N-\tau}} & =\frac{2t N}{\frac{2}{\delta} \parenv*{N- \sum_{i=0}^{t-\delta} \binom{n-\delta}{i} (\kp+\km)^{i+\delta}}} \\
& \leq \frac{2t N}{\frac{2}{\delta} \parenv*{N-N/(\kp+\km)^\delta }} \leq 2t\delta.
\end{align*}
\end{IEEEproof}

Now we can present our reconstruction process in  Algorithm~\ref{alg:reconstruction}.

\begin{algorithm}[ht]
  \caption{Reconstruction algorithm for $\km>0$}
  \label{alg:reconstruction}
  \begin{algorithmic}[1]
    \Statex \textbf{Input:} an $N$-set $\cY=\set{\vy_1,\vy_2,\ldots,\vy_N} \subseteq {\vx} +\BALL$ for some $\vx\in\cC$, and a threshold $\tau$
    \Statex \textbf{Output:} the codeword $\vx\in\cC$
    \State $\vz \gets $ the output of Algorithm~\ref{alg:majority} with $\cY$ and $\tau$ being the input
    \State $\cE \gets \set{i \in [1,n] ; \vz[i]=?}$
    \State $\cU \gets \set{\vu \in \Z^n; \vu[i]=\vz[i] \textup{ for all } i \not \in \cE \textup{ and } \vu[i] \in [\vy_1[i]-\kp, \vy_1[i]+\km] \textup{ for all } i \in \cE   }$
    \For{$\vu \in \cU$} 
    \State $\vx \gets \cD_{\cC}(\vu)$
    \If{$\cY \subseteq \vx+\BALL$}
    \State \Return{$\vx$}
    \EndIf
    \EndFor
  \end{algorithmic}
\end{algorithm}

\begin{theorem}Let $\cC \subseteq \Z^n$ be a code with the minimum distance $d_{\kp,\km}(\cC)=\delta$. Let $N$ and $\tau$ be defined as in \eqref{eq:decnumofread} and \eqref{eq:decthreshold}.
Let $\cY=\set{\vy_1,\vy_2,\ldots,\vy_{N}}$ be a set of $N$  vectors  coming from the same ball $\vx+\cB(n,t,\kp,0)$ for some codeword $\vx \in \cC$.
Then we can reconstruct $\vx$ by applying Algorithm~\ref{alg:reconstruction} on $\cY$ and $\tau$ with time complexity $O(Nn+C)$, where $C$ is the time complexity of the unique-decoding algorithm of $\cC$.
\end{theorem}

\begin{IEEEproof} Let $\cU$ be defined as in Algorithm~\ref{alg:reconstruction}.
According to Lemma~\ref{lm:boundonerror}, there is a vector $\vu\in \cU$ such that $\vu \in \vx+\cB(n,\delta-1 ,\kp,\km)$.  Thus we could apply the decoder $\cD_\cC$ of $\cC$ to each vector of $\cU$ to obtain a subset $\cS\subseteq \cC$ which contains $\vx$. Finally, since $\abs{\cY} =N> N(\cC;t,\kp,\km)$, the vector $\vx$ can be identified from  $\cS$ by checking each codeword $\vc \in \cS$ whether $\cY \subseteq \vc+\BALL$.

Let us analyze the time complexity of Algorithm~\ref{alg:reconstruction}. The complexity of Step 1 is $O(nN)$. According to Lemma~\ref{lm:bounderasure}, the decoding loop starting in Step 4 takes at most $(\kp+\km+1)^{2t\delta}$ rounds, which is independent of $n$. The complexity of checking the condition in Step 6 is also $O(nN)$. So the total time complexity of this algorithm is $O(nN+C)$.
\end{IEEEproof}

\section{List Decoding with Multiple Received Sequences}\label{sec:listreconstruction}

For a code $\cC\subseteq \Z^n$ with minimum distance $d_{\kp,\km}(\cC)=\delta$, denote $f\eqdef t-\delta+1$, that is, the number of excessive errors that $\cC$ cannot cope with. Let $\cY=\set{\vy_1,\vy_2,\ldots,\vy_N}$ be a subset of $\vx+\BALL$ for some $\vx \in \cC$.  In Section~\ref{sec:intersection} and Section~\ref{sec:algorithm}, we have shown that $\vx$ can be recovered  from $\cY$ if
\begin{equation}
\label{eq:reqsingle}
\begin{split}
 N  > & \begin{cases}
    (\kp)^{\delta} V_{\kp+1}(n-\delta,f-1), & \text{if $\km=0$,} \\
     (\kp+\km)^{2\delta} V_{\kp+\km+1}(n,f-1), & \text{otherwise.}
\end{cases}
\end{split}
\end{equation}

In this section, we introduce another degree of freedom into our setting, which is the decoder's ability to return a list of codewords instead of a single one. We show that by doing so, the decoder requires substantially fewer vectors from the channel, compared with~\eqref{eq:reqsingle}.  We shall modify the reconstruction algorithms in Section~\ref{sec:algorithm} to produce a list  of candidates $\cL$  which contains the transmitted codeword $\vx$. We first look at the case of $\km=0$.

\begin{theorem} Let $\cC \subseteq \Z^n$ be a code with the minimum distance $d_{\kp}(\cC)=\delta$.  Let $\vx \in \cC$ be a codeword and $\cY=\set{\vy_1,\vy_2,\ldots,\vy_N}$ be an $N$-subset of  $\vx+\cB(n,t,\kp,0)$.
If
\[N > (\kp)^{\delta+a} V_{\kp+1}(n-\delta-a,f-1-a),\]
where $0\leq a\leq f-1$, then we can decode to get a list $\cL\subseteq\cC$  containing $\vx$ with size
\[\abs{\cL} \leq  V_{\kp+1}(n,a).\]
Moreover, the time complexity of the decoding  is $O(nN+n^aC)$, where $C$ is the time complexity of the decoding algorithm of $\cC$.
\end{theorem}

\begin{IEEEproof}
Let $\vz\in\Z^n$ be defined by $\vz[i]=\min\{\vy_1[i],\vy_2[i],\ldots,\vy_{N}[i]\}$ for each $i \in [1,n]$. Since $N > (\kp)^{\delta+a} V_{\kp+1}(n-\delta-a,f-1-a)$, similarly to the proof of Theorem~\ref{thm:algorithm-km=0}, we can show that  $\vz \in \vx+\cB(n,\delta-1+a,\kp,0)$.
Then there is a vector $\vu \in \vz-\cB(n,a,\kp,0)$ such that $\vu \in \vx+\cB(n,\delta-1,\kp,0)$. Thus we may apply the decoding of $\cC$ on each vector of $\vz-\cB(n,a,\kp,0)$ to get the list $\cL$, i.e.,
\[ \cL\eqdef \set*{ \cD_{\cC}(\vu) ; \vu\in\vz-\cB(n,a,\kp,0)}.\]
It also follows that $\abs{\cL}\leq  V_{\kp+1}(n,a)$. The claimed complexity follows from the fact that $V_{\kp+1}(n,a)=O(n^a)$.
\end{IEEEproof}

Now we study the case of $\km >0$. Let $0\leq a\leq f-1=t-\delta$. Assume that
\begin{equation}\label{eq:listdecnumofread}
 N\eqdef  (\kp+\km)^{\delta+a+1} V_{\kp+\km+1}(n-\delta-a,f-1-a)+1.
 \end{equation}
 Denote
 \begin{equation}\label{eq:listdecthreshold}
 \tau  \eqdef \parenv*{1-\frac{2}{\delta+a}}N + \frac{2}{\delta+a} \sum_{i=0}^{t-\delta-a} \binom{n-\delta-a}{i} (\kp+\km)^{i+\delta+a}.
 \end{equation}

Given an $N$-set  $\cY=\set{\vy_1,\vy_2,\ldots,\vy_N} \subseteq \vx+\BALL$ for some vector $\vx$, we first apply Algorithm~\ref{alg:majority} with threshold $\tau$ on it to obtain an estimate $\vz$ of $\vx$.
Similar to Lemma~\ref{lm:boundonerror} and Lemma~\ref{lm:bounderasure}, we have the following result on $\vz$. The proofs are the same as those in Section~\ref{sec:algorithm} and we omit here.

\begin{lemma}\label{lm:boundonerror-list}
Let $\cY\subseteq \vx+\BALL$ be an $N$-set, $\vx\in\Z^n$, and let $\vz$ be the output of Algorithm~\ref{alg:majority} when run on $\cY$ with $\tau$ from~\eqref{eq:listdecthreshold}. Then $\vz$ contains at most $\delta+a-1$ errors compared with $\vx$, and at most $2t(\delta+a)$ erasures.
\end{lemma}
\begin{IEEEproof}
The proof is the same as those of Lemma~\ref{lm:boundonerror} and Lemma~\ref{lm:bounderasure}.
\end{IEEEproof}

Our list decoding algorithm is presented in  Algorithm~\ref{alg:listreconstruction}.

\begin{algorithm}[ht]
  \caption{List decoding algorithm}
  \label{alg:listreconstruction}
  \begin{algorithmic}[1]
    \Statex \textbf{Input:} an $N$-set $\cY=\set{\vy_1,\vy_2,\ldots,\vy_N} \subseteq \vx+\BALL$ for some $\vx\in\cC$, and a threshold $\tau$
    \Statex \textbf{Output:} a set $\cL \subseteq \cC$ such that $\vx \in \cL$
    \State $\vz \gets $ the output of Algorithm~\ref{alg:majority} with $\cY$ and $\tau$ being the input
    \State $\cE \gets \set{i \in [1,n] ; \vz[i]=?}$
    \State $\cU \gets \set{\vu \in \Z^n; \vu[i]=\vz[i] \textup{ for all } i \not \in \cE \textup{ and } \vu[i] \in [\vy_1[i]-\kp, \vy_1[i]+\km] \textup{ for all } i \in \cE   }$
    \State $\cV \gets \bigcup_{\vu \in \cU} \parenv*{ \vu -\cB(n,a,\kp,\km)}$
    \State $\cL \gets \set{ \cD_\cC(\vv) ; \vv\in\cV }$
    \State \Return{$\cL$}
  \end{algorithmic}
\end{algorithm}

\begin{theorem}\label{thm:listalgorithm} Let $\cC \subseteq \Z^n$ be a code with the minimum distance $d_{\kp,\km}(\cC)=\delta$. Let $N$ and $\tau$ be defined as in \eqref{eq:listdecnumofread} and \eqref{eq:listdecthreshold}.
Let $\cY=\set{\vy_1,\vy_2,\ldots,\vy_{N}}$ be a set of $N$  vectors contained in the same ball $\vx+\cB(n,t,\kp,0)$ for some codeword $\vx \in \cC$.
Then we can decode a list $\cL$ containing $\vx$ by applying Algorithm~\ref{alg:listreconstruction} on $\cY$ and $\tau$, where
 \[\abs{\cL} \leq (\kp+\km+1)^{2t(\delta+a)} V_{\kp+\km+1}(n,a).\]
The  time complexity is $O(Nn+n^aC)$, where $C$ is the time complexity of the decoding algorithm of $\cC$.
\end{theorem}

\begin{IEEEproof} Let $\cU$ be defined as in Algorithm~\ref{alg:listreconstruction}.
According to Lemma~\ref{lm:boundonerror-list}, there is a vector $\vu\in \cU$ such that $\vu \in \vx+\cB(n,\delta+a-1 ,\kp,\km)$.
Since $\cV = \bigcup_{\vu \in \cU} \parenv*{ \vu -\cB(n,a,\kp,\km)}$, we can find a vector $\vv \in \cV$ such that $\vu \in \vx+\cB(n,\delta-1 ,\kp,\km)$. Thus we can apply the decoder $\cD_\cC$ of $\cC$ to each vector of $\cV$ to obtain a subset $\cL\subseteq\cC$ which contains $\vx$. Moreover, the size of list
\[\abs{\cL} \leq  \abs{\cV} \leq  \abs{\cU}V_{\kp+\km+1}(n,a)\leq(\kp+\km+1)^{2t(\delta+a)} V_{\kp+\km+1}(n,a),\]
 where the last inequality holds as Lemma~\ref{lm:boundonerror-list} implies that $\abs{\cU} \leq (\kp+\km+1)^{2t(\delta+a)}$.

Let us analyze the time complexity of Algorithm~\ref{alg:listreconstruction}. The complexity of Step 1 is $O(nN)$. The upper bound on $\abs{\cL}$ also bounds the number of times we run $\cD_\cC$, hence, we use the unique-decoder for $\cC$ at most $O(n^a)$ times. Thus, the total time complexity of this algorithm is $O(nN+n^a C)$.
\end{IEEEproof}

In \cite{JunLaiLeh21},  the  trade-off between the size of the minimum list  and the number of different received noisy sequences   has been analyzed  for substitutions. Modifying the approach therein, we could give another list-decoding algorithm which reduces simultaneously the value of $N$ and the size of $\cL$ in Theorem~\ref{thm:listalgorithm}, at the cost of  the time complexity.
We require the following $q$-ary Sauer-Shelah lemma.

\begin{lemma}[\cite{GuruHasKop10}]\label{lm:qSauerShelah}
For all integers $q,n,c$ with $c\leq n$, for any set $\cS \subseteq [0,q-1]^n$, if $\abs{\cS} > V_{q}(n,c-1)$, then there exists some set of coordinates $U \subseteq [1,n]$ with $\abs{U}=c$ such that for every $\vu \in [0,q-1]^U$, there exists some $\vv \in\cS$ such that $\vu$ and $\vv|_U$ differ in every position.
\end{lemma}

\begin{remark}
The original condition in \cite{GuruHasKop10} is $\abs{\cS} > 2((q-1)n)^{c-1}$. However, even if we replace it with $\abs{\cS} > V_{q}(n,c-1)$, the proof still works and so the conclusion still holds.
\end{remark}

\begin{theorem}\label{thm:listalgorithm-3}
Let $\cC \subseteq \Z^n$ be a code with $d_{\kp,\km}(\cC) = \delta$. Let $\vx \in \cC$ be a codeword and $\cY=\set{\vy_1,\vy_2,\ldots,\vy_N}$ be a subset of  $\vx+\BALL$ of size $N$.
 If  $N > V_{\kp+\km+1}(n,f-1-a)$ where $0\leq a \leq f-1$, then we can decode to get a list $\cL$  containing $\vx$ with size
\[\abs{\cL} \leq (\kp+\km+1)^{2(f-a)}  V_{\kp+\km+1}(n-f+a,a).\]
\end{theorem}

\begin{IEEEproof}
For each $i \in [1,n]$,
define
\begin{align*}
m_i &\eqdef\min\set*{\vy[i]; \vy \in \cY}, \\
M_i & \eqdef\max\set*{\vy[i];\vy\in \cY}, \\
K_i &\eqdef [ \min\set*{m_i, M_i-\kp}, \max\set*{M_i, m_i+\km}].
\end{align*}
Then $\abs{K_i} \leq \kp+\km+1$ for all $i \in [1,n]$. Furthermore, we have
\[\vx \in K_1\times K_2\times \cdots \times K_n\] and
\[\vy_\ell\in K_1\times K_2\times \cdots \times K_n\] for every $\ell \in [1,N]$. Since $N >V_{\kp+\km+1}(n,f-1-a)$, according to Lemma~\ref{lm:qSauerShelah}, there is a subset $U\subseteq [1,n]$ of size $f-a$ and a vector $\vy_{\ell_0} \in \cY$ such that $\vx$ differs from $\vy_{\ell_0}$ in every position of $U$.

Let $\cY'$ be a minimal subset of $\cY$ such that for every  $\vy \in \cY$ there is a vector $\vy'\in \cY'$ with $\vy$ and $\vy'$ being the same on $U$. Then
$\abs{\cY'} \leq (\kp+\km+1)^{\abs{U}}$, and there  is also a vector $\vy^*\in\cY'$ such that $\vx$ differs from $\vy^*$ in every position of $U$.
Let
\[\cD \eqdef \bigcup_{\vy'\in \cY'} \set*{\vz\in \vy'- \cB(n,f,\kp,\km);   \vz[i]\neq \vy'[i] \textup{ for every } i \in U }.\]
Since $\vy'\in \vx+\BALL$ for all $\vy'\in \cY'$ and $f=t-\delta+1$,  there is a vector $\vz^*\in \cD$ such that $\vz^*$ and $\vx$ agree in every position of $U$ and $\vz^* \in \vx+\cB(n,\delta-1,\kp,\km)$.
Let
\[\cL = \set{\cD_{\cC}(\vz); \vz\in \cD  }, \]
where $\cD_{\cC}$ is the decoder of $\cC$.
Then $\vx \in \cL$ and the size
\begin{align*}
\abs{\cL} &\leq \abs{\cD} \leq \abs{\cY'}(\kp+\km)^{\abs{U}} V_{\kp+\km+1}(n-\abs{U}, f-\abs{U}) \\
&<(\kp+\km+1)^{2(f-a)} V_{\kp+\km+1}(n-f+a, a).
\end{align*}
\end{IEEEproof}
To the best of  our knowledge, there is no efficient algorithm to identify the subset $\cU$ in the  Sauer-Shelah lemma. A brute-force algorithm  requires $O(n^{f-a}N)$ comparisons. Thus the total time complexity of the decoding is $O(n^{f-a}N+n^aC)$, whereas the complexity of Algorithm~\ref{alg:listreconstruction} is $O(nN+n^aC)$.

Theorem~\ref{thm:listalgorithm-3} requires at least $V_{\kp+\km+1}(n,f-1-a)+1$ distinct received sequences to obtain a list of size $O(n^a)$. A natural question that arises is whether this requirement is tight? The following lemma, modified from~\cite[Lemma 32]{JunLaiLeh21}, shows that it is almost tight. The lemma shows that if $N \leq V_{\kp+\km}(n,f-1-a)$, there is a code $\cC\subseteq\Z^n$, and a list $\cL\subseteq\cC$ of size $\Omega(n^{a+1})$ such that $\cY \subseteq \bigcap_{\vu \in \cL}(\vu+\BALL)$, i.e., $\cY$ is in the intersection of too many balls around codewords.

\begin{lemma}Assume that $(\kp,\km)\neq (1,0)$. Let $N \leq  V_{\kp+\km}(n,f-a)$, where $0\leq a \leq f$ and
$n\geq 2e+a$. Then there is a set of vectors $\cY \subseteq \Z^n$ with $\abs{\cY}=N$ and a code $\cC \subseteq \Z^n$ of size
 \[\abs{\cC} \geq \frac{n^a}{ (e+a)^a \sum_{i=0}^e \binom{e+a}{i} },\]
such that
\begin{itemize}
\item  $\cC$  can correct $e$ $(\kp,\km)$-limited-magnitude errors; and
\item $\cY \subseteq \bigcap_{\vx \in \cC} \parenv*{ \vx+\BALL}$.
\end{itemize}
\end{lemma}

\begin{IEEEproof}
Let $\cS =\set{\vv \in [-\km,\kp-1]^n; \wt(\vv) \leq f-a}$ and $\cY$ be an arbitrary subset of $\cS$ with $\abs{\cY}=N$. Let $\cC\subseteq \set{-1,0}^n$ be a binary code  with minimum Hamming distance $2e+2$ and constant weight $e+a$. The lower bound on the size of $\cC$ comes from the Gilbert-Varshamov bound.
Since $\cC$ can correct $e$ substitutions, it also can correct the same number of  $(\kp,\km)$-limited-magnitude errors.
Noting that $e+a+f-a=t$, we have that $\cY \subseteq \bigcap_{\vx \in \cC} \parenv*{\vx+\BALL}.$
\end{IEEEproof}

\section{Reconstruction for Uniform Tandem Duplications}\label{sec:tandemdup}
In this section, we show that our reconstruction algorithm for $(\kp,0)$-limited-magnitude errors (Algorithm~\ref{alg:minimum}) can also be used for \emph{tandem duplications}, which create a copy of a block of the sequence and insert it in a tandem manner, i.e., next to the original. 

The design of reconstruction codes against $t$ tandem duplications of the same length $k$ was studied in \cite{YehSch20}. 
Such a  code  could be decomposed into a family of subcodes $\cC_{\vx}$'s, so that  the codewords from the same subcode shares the same \emph{root} $\vx$. This vector $\vx$ could be computed from the codeword in linear time and is robust against any number of tandem duplications of the same length $k$. Thus,
for any two codewords $\vu$ and $\vu'$ from different subcodes, they are always distinguishable from each other no matter how many tandem duplications of length $k$ affect them. Thus, in order to study the decoding/reconstruction problem for tandem duplications, it suffices to consider the corresponding problem for the code of  $\cC_{\vx}$. 

Under certain mapping $\psi_{\vx}$, each subcode $\cC_{\vx}$ can be embedded into the simplex $\Delta_{r(\vx)}^{m(\vx)+1}$, where $m(\vx)$ and $r(\vx)$ are some integers determined by $\vx$, and 
\[\Delta_r^m\eqdef\set*{\vx=(x_1,x_2,\ldots,x_m+1) \in \N^{m+1}; \sum_{i=1}^{m+1}x_i=r}. \]
A tandem duplication of length $k$ in $\vu$ corresponds to an addition of a unit vector $\ve_j \in \N^{m(x)+1}$ to $\psi_{\vx}(\vu)$. Thus, in order to design a reconstruction code with maximum intersection less than $N$, it is required that for any two distinct codewords $\vu,\vu' \in \cC_{\vx}$, it always holds that $d_{\ell_1}(\psi_{\vx}(\vu),\psi_{\vx}(\vu')) \geq 2\delta$,  where $\delta$ is the minimum integer such that $\binom{t-\delta+m(\vx)}{m(\vx)}< N$.
For more details on the code construction and its relation
to constant-weight integer codes in the Manhattan metric, the reader may refer to \cite[Section~III]{YehSch20}. Note that the notation $N$ in this section represents the number of reads,  while the same notation in \cite[Section~III]{YehSch20} represents the designed size of maximum intersection.

For a vector $\vx\in \N^{m+1}$, let  
\[\cB_t^+(\vx)\eqdef \set*{\vy\in \N^{m+1}; y_i\geq x_i \textup{ for all } 1\leq i \leq m+1, \textup{ and } \sum_{i=1}^{m+1} (y_i-x_i) \leq t}.\] 
Then the reconstruction problem for the  code in \cite[Section~III]{YehSch20} can be reduced as follows: Given a code $\cC \subseteq \Delta_r^{m+1}$ and a set of vectors $\cY=\set{\vy_1,\vy_2,\ldots,\vy_N}$ such that $\cY \subseteq \cB_t^+(\vx)$ for some $\vx \in \cC$, we would like to reconstruct $\vx$ from $\cY$. To this end, we use the same decoding process as Algorithm~\ref{alg:minimum}.
Let $\vz=(z_1,z_2,\ldots,z_{m+1})$ where $z_i=\min\{\vy_1[i],\vz_y[i],\ldots,\vy_{N}[i]\}$ for each $i \in [1,m+1]$.
It is easy to see that $\vx[i] \leq z_i$ for each $i \in [1,m+1]$. Thus, $\vz \in \cB_r^+(\vx)$ for some  integer $r$. If $r \geq \delta$, then there exist some positions $i_1,i_2,\ldots,i_{\tau}$ and some positive integers $\delta_1,\delta_2,\ldots,\delta_\tau$ such that $\sum_{j=1}^\tau \delta_j =\delta$ and $\vy_\ell[i_j]-\vx[i_j] \geq \delta_{j}$ for each $j \in [1,\tau]$ and $\ell\in[1,N]$. However, since $N> \binom{m+t-\delta}{m}$, it is impossible. Hence, $r< \delta$ and we may run the decoding algorithm of $\cC$ on $\vz$ to recover $\vx$.

\bibliographystyle{IEEEtranS}
\bibliography{allbib}

\end{document}